\documentclass{ws-p8-50x6-00}
\newcommand{\xbf}[1]{\mbox{\boldmath $ #1 $}}
\begin{document}

\title{Neutron charge form factor and  \\
quadrupole deformation of the nucleon\footnote{NSTAR 2001, 
Proceedings of the Workshop on the Physics of  Excited Nucleons, 
Mainz, Germany, 7-10 March 2001,
Eds. D. Drechsel and  L. Tiator, World Scientific, Singapore, 2001, pg. 229}
}

\author{A. J. BUCHMANN} 
\address{Institut f\"{u}r Theoretische Physik, Eberhard Karls Universit\"{a}t
T\"ubingen, \\ 
Auf der Morgenstelle 14, D-72076 T\"ubingen, Germany\\
E-mail: alfons.buchmann@.uni-tuebingen.de}

\maketitle

\abstracts{A quark model relation between the neutron charge form factor 
and the $N \to \Delta$ charge quadrupole 
form factor is used to predict the $C2/M1$ 
ratio in the $N \to \Delta$ transition from the elastic neutron 
form factor data. Excellent agreement with 
the electro-pionproduction data is found, indicating the validity
of the suggested relation. The implication of the negative $C2/M1$ ratio 
for the intrinsic deformation of the nucleon is discussed.}

The geometrical shape of the proton can be determined 
from its {\it intrinsic} quadrupole moment 
\cite{Boh75} 
\be 
Q^p_0=\int \!d^3r\, \rho^p({\bf r})\, (3 z^2 - r^2), 
\ee
where $\rho^p({\bf r})$ is the not necessarily spherically symmetric 
charge density of the proton. If $Q_0^p > 0$, the proton is prolate 
(cigar-shaped);
if $Q_0^p < 0$, it is oblate (pancake-shaped); and if 
$Q_0^p=0$, the charge density inside the proton is spherically symmetric.
The intrinsic quadrupole moment,
which is defined with respect to the body-fixed frame, must be distinguished
from the {\it spectroscopic} quadrupole moment 
measured in the laboratory. 
The latter is zero due to angular momentum selection rules.
$Q^p_0$ can be inferred by measuring electromagnetic quadrupole transitions 
between the nucleon ground and its low-lying excited states, or 
by measuring the quadrupole moment of an excited state, e.g. 
the $\Delta(1232)$ with $J > 1/2$.  

Another possible way to obtain information on the shape 
of the nucleon has recently been suggested~\cite{Buc97,Buc00,Hen01}. 
In the constituent quark model
with two-body exchange currents a connection between the neutron 
charge form factor $G_{C0}^n({\bf q}^2)$ and the $N \to \Delta$ 
charge quadrupole form factor $G_{C2}^{p \to \Delta^+}({\bf q}^2)$ 
has been found~\cite{Buc00}
\be
\label{rel1}
G_{C2}^{p \to \Delta^+}\!({\bf q}^2)= -\frac{3 \sqrt{2}}{{\bf q}^2} 
\, G_{C0}^n({\bf q}^2),
\ee
where ${\bf q}$ is the three-momentum transfer of the virtual photon.
Together with the known SU(6) relation 
$G_{M1}^{p \to \Delta}({\bf q}^2)\!=\!-\sqrt{2} G_{M1}^n({\bf q}^2)$, 
which remains valid after adding exchange currents~\cite{Buc00}, this
provides a determination of the C2/M1 ratio through 
the {\it elastic} charge and magnetic 
neutron form factors. The $C2/M1$ ratio obtained in this way agrees well with C2/M1 data from 
electro-pionproduction experiments (see Fig.1)~\cite{Gra01}. 
By comparing the low ${\bf q}$ expansion of the left and right hand side of Eq.(\ref{rel1}),
the $N\to \Delta$ quadrupole moment $Q_{p\to \Delta}$ can be expressed 
in terms of the known neutron charge radius $r_n^2$. Likewise, the    
$N\to \Delta$ quadrupole transition radius $r^2_{p\to \Delta^+}$ 
is obtained from the fourth moment $r_n^4$ of the neutron's charge density $\rho^n(r)$
\be 
\label{rel2}
 Q_{p \to \Delta^+} = \frac{1}{\sqrt{2}}\, r_n^2 \, ,
\quad r_{p \to \Delta^+}^2= \frac{3}{10}\frac{r_n^4}{r_n^2}.  
\ee
Experimentally~\cite{Gra01}  one finds $r_n^2=-0.113(3)$ fm$^2$ 
and $r_n^4\!=\!-0.32(8)$ fm$^4$, from which we predict 
$Q_{p \to \Delta^+}=-0.080(2)$ fm$^2$ 
and $r^2_{p \to \Delta^+}\!\!=\!\!-0.84(21)$ fm$^2$. 

\begin{center}
\begin{figure}[htb]
\begin{center}
\epsfxsize 11.5 true cm
\epsfysize 5.0 true cm
\epsfbox{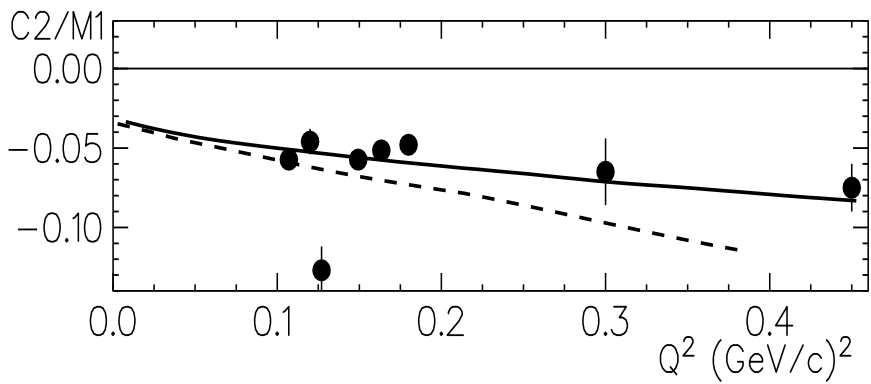}
\vspace{-0.3 cm}
\caption{\label{fig:ratio}
The C2/M1 ratio in the electromagnetic $N \to \Delta$ transition: 
(i) explicit quark model calculation(dashed curve)~\protect\cite{Buc00};
(ii) as obtained from the {\it elastic} neutron form factor data 
according to Eq.(\ref{rel1})(solid curve)~\protect\cite{Gra01}; 
(iii) experimental electro-pionproduction data~\protect\cite{Got01}.}
\end{center}
\end{figure}
\end{center}

Quite generally, the baryon's charge density consists of a 
sum of one-, two-, and  three-quark pieces: 
$
\rho({\bf q})= \rho_{[1]}({\bf q}) +
\rho_{[2]}({\bf q}) + \rho_{[3]}({\bf q}).
$
The two- and three-quark terms describe the nonvalence 
quark degrees of freedom (e.g. $q \bar q$ pairs) seen by the 
electromagnetic probe. In deriving Eq.(\ref{rel1}) we have made the following 
assumptions: (i) the baryon wave functions with orbital angular momentum $L=0$ 
contain only valence quark degrees of freedom;
(ii) for $G_{C0}^n$ and $G_{C2}^{p \to \Delta^+}$ one- and three-body 
operators are suppressed in comparison to the two-body term $\rho_{[2]}$.
Assumption (i) is not a restriction provided the nonvalence degrees of 
freedom are included in the form of many-body operators~\cite{morp},
and assumption (ii) is supported by several investigations~\cite{Buc97,Hen01,lebed}.
Eq.(\ref{rel1}) is then a result of the dominance of $\rho_{[2]}$ 
for both observables, the spin-flavor structure of $\rho_{[2]}$, and the SU(6) spin-flavor 
symmetry of the $N$ and $\Delta$ wave functions. 
\begin{figure}[htb]
$$\mbox{
\epsfxsize 12.0 true cm
\epsfysize 6.0 true cm
\setbox0= \vbox{
\hbox{ 
\epsfbox{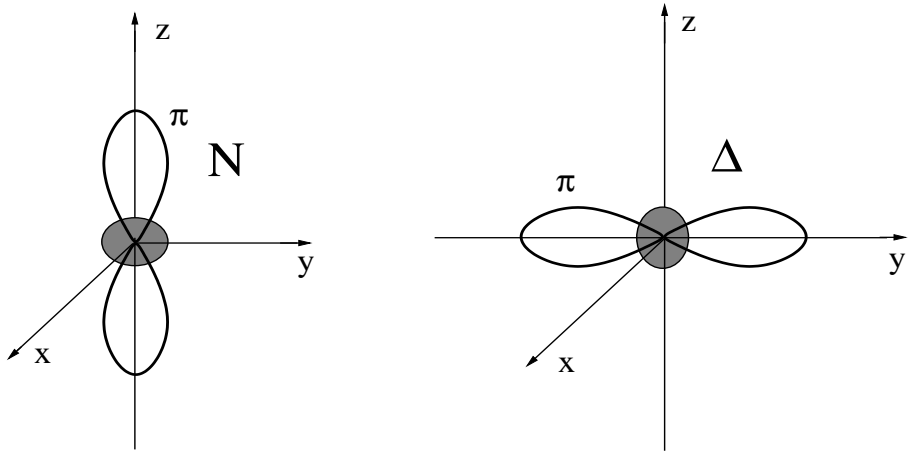}
} 
} 
\box0
} $$
\vspace{-0.25cm}
\caption[Pion cloud ]{Intrinsic quadrupole deformation of the nucleon (left)
and $\Delta$ (right) in the pion cloud model. In the $N$  
the $p$-wave pion cloud is concentrated along the polar (symmetry) axis, 
with maximum probability of finding the pion at the poles. 
This leads to a prolate deformation. In the $\Delta$, the pion cloud is 
concentrated in the equatorial plane producing an oblate intrinsic 
deformation. }
\label{fig:pcm}
\end{figure}

This can be seen as follows. A multipole 
expansion of, e.g., the gluon exchange charge operator 
$\rho_{[2]}$ up to quadrupole terms 
gives with ${\bf q}= q{\bf e}_z$ 
the following decomposition in spin-isospin space
\bea
\label{eq:structure}
\rho_{[2]} &= & -B \sum_{i < j}
 e_i \biggl (  2\xbf{\sigma}_i \cdot \xbf{\sigma}_j
-( 3 \xbf{\sigma}_{i\, z} \, \xbf{\sigma}_{j \, z} 
-\xbf{\sigma}_i \cdot \xbf{\sigma}_j ) \biggr ) + (i \leftrightarrow j) 
\eea
where $\xbf{\sigma}_{i\, z}$ is the $z$ 
component of the spin operator of quark $i$, and $e_i$ is the quark charge.
$B$ contains the orbital and color part common to the spin-scalar (C0) 
and spin-tensor (C2) part of the operator. 
Note that there is a {\it fixed} relative strength between the C0 term 
and the C2 term. Evaluating $\rho_{[2]}$ between SU(6) spin-flavor 
$N$ and $\Delta$ wave functions, one obtains $r_n^2=4B$ 
and $\sqrt{2} Q_{p \to \Delta^+}=4B$. 
This leads to the first equation in Eq.(\ref{rel2}). A generalization 
of this derivation to finite momentum transfers is straightforward and 
leads to Eq.(\ref{rel1}).

In order to calculate $Q_0^p$ from the observable $Q_{p \to \Delta^+}$ 
we need a model. We have calculated $Q_0^p$ using three different nucleon
models~\cite{Hen01}.
In the quark model we find that Eq.(\ref{rel2}) implies
\be
\label{rel3}
Q_0^p=-Q_0^{\Delta^+}=-r_n^2 \, ,
\ee 
i.e., a prolate intrinsic deformation of the proton and 
an oblate intrinsic deformation of the $\Delta^+$. We also see that the 
neutron charge radius $r_n^2$ and the quadrupole deformation of the nucleon 
are intimately related phenomena, which reflect the $q \bar q$ 
degrees of freedom in the nucleon.

Also in the pion cloud model~\cite{Hen62} 
a relation between $r_n^2$ and 
$Q_{p \to \Delta^+}$ and between the intrinsic quadrupole 
moments of the $N$ and $\Delta$ is obtained~\cite{Hen01} 
\be 
Q_{p \to \Delta^+}=Q_{\Delta^+}=r_n^2, \qquad
Q_0^{p}= -Q_0^{\Delta^+} = -r_n^2.
\ee
Even though the valence quark core of the nucleon 
may have a small oblate deformation
(as one would obtain from the small D-state admixture of the valence quarks), 
the major contribution to the intrinsic quadrupole moment comes from 
the p-wave coupling 
of the pion cloud to the valence quark core. From angular
momentum coupling the pion cloud in the proton is oriented along the polar 
axis 
(see Fig.2). This leads to an overall prolate deformation of the nucleon.
Similarly, for the $\Delta$ with spin 3/2, angular momentum conservation 
implies that the pion cloud lies in the equatorial plane
characteristic of an oblate intrinsic deformation. In addition, we have 
calculated $Q_0^p$ and $Q_0^{\Delta^+}$ in the Bohr-Mottelson collective 
model~\cite{Hen01}, with the same qualitative results, namely 
a prolate shape of the $N$ and an oblate shape of the $\Delta$.

\end{document}